\documentclass[preprint,3p,twocolumn]{elsarticle}
\usepackage[utf8]{inputenc}
\usepackage{leftindex}
\usepackage{xcolor}
\usepackage{amsfonts}
\usepackage{graphicx}
\usepackage{booktabs}
\usepackage{multirow}

\allowdisplaybreaks

\begin{document}
\begin{frontmatter}

\title{Supporting Data Center Grid Integration via Co-Optimized Battery Energy Storage Sizing and Capacity Allocation}
\tnotetext[t1]{This research was supported by the Energy System Co-Design with Multiple Objectives and Power Electronics (E-COMP) Initiative, under the Laboratory Directed Research and Development (LDRD) Program at Pacific Northwest National Laboratory (PNNL). PNNL is a multi-program national laboratory operated for the U.S. Department of Energy (DOE) by Battelle Memorial Institute under Contract No. DE-AC05-76RL01830.}

\author[1]{Wei Wang}
\ead{w.wang@pnnl.gov}
\author[1]{Himanshu Sharma\corref{cor1}}
\ead{himanshu.sharma@pnnl.gov}
\author[1]{Thiagarajan Ramachandran}
\ead{thiagarajan.ramachandran@pnnl.gov}
\cortext[cor1]{Corresponding author}
\affiliation[1]{organization={Pacific Northwest National Laboratory},
city={Richland, WA 99352},
country={United States}}

\begin{abstract}
Interconnecting large data centers can be delayed by grid requirements on demand flexibility, ramping limits, and ride-through performance. This paper proposes an optimization-based framework to size a Battery Energy Storage System (BESS) that accelerates data center grid integration by providing BESS capacity required to meet selected flexibility and grid-support requirements. The model co-optimizes BESS power rating and energy capacity while allocating capability across three services: (i) demand response over a specified duration, (ii) mitigation of high-magnitude, high-frequency load fluctuations through reserved active power with zero net state-of-charge change, and (iii) voltage support via reactive power capability at the point of common coupling. Minute-level operational scheduling enforces state-of-charge feasibility and load ramp constraints, and incorporates a controllable queue of delay-insensitive computing tasks to enhance flexibility. A case study using a real data center load profile demonstrates the proposed method and quantifies the BESS power-energy requirements and operational trade-offs needed to satisfy the multi-service constraints, providing practical guidance for interconnection planning and deployment. 
\end{abstract}



\end{frontmatter}

\section*{Notations}


\noindent
\textbf{Coefficients}
\begin{description}
\setlength{\labelwidth}{1cm}
\setlength{\leftmargin}{1cm}
\setlength{\itemindent}{0pt}
\setlength{\labelsep}{0.5em}
  \item[$C^E$] BESS unit energy capacity cost
  \item[$C^P$] BESS unit rated power cost
  \item[$d^\mathrm{IT}_t$] Total data center load at time $t$
  \item[$\hat d^\mathrm{IT}_t$] Time sensitive load at time $t$
  \item[$D^\mathrm{dr}_t$] Demand response request signal
  \item[$\overline E^\mathrm{BESS}$] BESS maximum energy capacity
  \item[$F^\mathrm{vs}_t$] PCC voltag sag signal at time $t$
  \item[$\overline P^\mathrm{BESS}$] BESS maximum rated power
  \item[$P^\mathrm{ramp}$] Ramp limit
  \item[$\overline P^\mathrm{DC}$] Data center upper load limit
  \item[$\overline P^\mathrm{Q}$] Maximum unprocessed load in the queue
  \item[$P^\mathrm{R}$] Data center rated load
  \item[$\mathrm{PUE}$] Power usage effectiveness
  \item[$r$] Resistance of main transmission line
  \item[$T^\mathrm{dr}$] Time duration for providing demand response
  \item[$T^\mathrm{L}$] Duration of one time step
  \item[$V^0_t$] Voltage of PCC before fault at time $t$
  \item[$V^\mathrm{ref}$] Reference voltage of root node
  \item[$\Delta V$] Allowed voltage deviation of root node
  \item[$x$] Reactance of main transmission line
  \item[$\alpha$] Fraction of rated load for demand response
  \item[$\beta$] Fraction of BESS rating for fast fluctuation smoothing
  \item[$\gamma$] Voltage drop of PCC during fault
  \item[$\eta^\mathrm{ch}$] Charge efficiency of BESS
  \item[$\eta^\mathrm{dis}$] Discharge efficiency of BESS
\end{description}

\noindent
\textbf{Decision variables}
\begin{description}
\setlength{\labelwidth}{1cm}
\setlength{\leftmargin}{0pt}
\setlength{\itemindent}{0pt}
\setlength{\labelsep}{0.5em}
  \item[$E^\mathrm{dr}$] Reserved energy capacity for demand response
  \item[$p^\mathrm{q}_t$] Unprocessed IT load in queue at time $t$
  \item[$\overline P^\mathrm{BESS}$] BESS maximum rated power
  \item[$p^\mathrm{IT}_t$] Processed IT load at time $t$
  \item[$p^\mathrm{Cool}_t$] Power consumption of cooling system at time $t$
  \item[$p^\mathrm{DC}_t$] Data center total load at time $t$
  \item[$p_{t}^\mathrm{ch_{op}}$] BESS charged power for DC operation at time $t$
  \item[$p_{t}^\mathrm{dis_{op}}$] BESS discharged power for DC operation at time $t$
  \item[$P^\mathrm{fs}$] BESS power for providing fast load fluctuation smoothing
  \item[$p^\mathrm{ch_{dr}}$] BESS charged power for providing flexibility
  \item[$p^\mathrm{dis_{dr}}$] BESS discharged power for providing flexibility
  \item[$P^\mathrm{Batt}$] BESS power for DC operation
  \item[$p^l_t$] Real power through main transmission line at time $t$
  \item[$q^l_t$] Reactive power through main transmission line at time $t$
  \item[$q_{t}^\mathrm{vs}$] Reactive power for voltage support
  \item[$SoC_{t}$] State of charge of BESS at time $t$
  \item[$V^\mathrm{DC}_t$] Voltage of DC root node after fault at time $t$
\end{description}

\section{Introduction}
The United States has seen an incredible increasing investment in artificial intelligence technologies, which has created great demand on energy supply. The 2024 Report on U.S. Data Center Energy Use finds that data centers consumed about 4.4\% of total U.S. electricity in 2023 and are expected to consume approximately 6.7 to 12\% of total U.S. electricity by 2028. The report indicates that total data center electricity usage climbed from 58 TWh in 2014 to 176 TWh in 2023 and estimates an increase between 325 to 580 TWh by 2028~\cite{shehabi2024datacenter}.

The rapid growth of large-scale data centers (DCs) is creating significant new electricity demand and increasing pressure on power system infrastructure. In many regions, it may take years for a project to get online~\cite{datacenterknowledge2025ai}. Large data centers can exhibit high and rapidly varying power demand due to changing computational workloads, while utilities and grid operators increasingly expect new loads to be more grid-friendly and capable of supporting reliable system operation. Data centers with demand flexibility, controllable load ramping behavior, and the ability to remain connected during voltage sags caused by main grid faults can convert such large load from purely grid burdens to grid assets~\cite{chen2020internet,xie2025enhancing,takci2025data}. As a result, proactively providing grid supports has become an important design consideration for new data center facilities.

Battery energy storage systems (BESS) offer a practical way to accelerate data center grid integration by enabling multiple flexibility and grid-support functions. With a properly sized BESS, a data center can reduce its net load during demand response events, limit short-term and medium-term power fluctuations, satisfy load ramping requirements, and maintain service continuity under upstream voltage disturbances through voltage ride-through (VRT) support. 

Mamun et al. proposed a multi-objective model that maximizes dollar savings through peak shaving while minimizing battery degradation~\cite{mamun2016multi}. Their results showed a significant Pareto tradeoff between the two objectives. Dabbagh et al. investigated the joint use of energy storage and workload shifting to reduce data-center peak demand and demonstrated that BESS can effectively mitigate short-duration demand spikes while maintaining service quality~\cite{dabbagh2017shaving}. Shi et al. further demonstrated that BESSs can reduce electricity costs by simultaneously performing peak shaving and participating in ancillary service markets such as frequency regulation~\cite{shi2016leveraging}. Cupelli et al. proposed control strategies that enable data centers to participate in DR programs by leveraging their heating, ventilation, and air conditioning unit, delay-tolerant information technology workload and battery storage system~\cite{cupelli2018data}. Azizi et al. proposed a grid-forming BESS framework that can significantly attenuate fast power fluctuations, improve voltage and power quality at the grid interface, and enhance overall data center reliability~\cite{azizi2026strengthening}. Xie et al. designed voltage controllers for data centers' internal power distribution network to enhance their low voltage ride-through capability, avoiding the tripping of data centers during main grid faults~\cite{xie2025enhancing}. Liu et al. developed a co-optimization framework that jointly schedules computing workloads and BESS dispatch under utility-imposed peak-load and ramp-rate constraints~\cite{liu2026watts}. Thompson et al. proposed a BESS sizing framework for data-centers to reduce microgrid cost and enhancing reliability~\cite{thompson2016optimization}. Cupelli et al. proposed a methodology for optimally sizing BESSs in data centers providing frequency containment reserve services while considering battery operational constraints and energy management strategies~\cite{cupelli2017optimal}.

The works mentioned above either focus on controlling/dispatching existing BESS of data centers or optimally sizing the BESS for specific usage. Nowadays, BESS is able to provide several capabilities for data centers but these services impose different technical requirements on the BESS. For example, demand response is primarily constrained by both power and sustained energy availability, fluctuation smoothing requires substantial power capacity, and VRT support requires additional apparent power capability for reactive power injection. Therefore, determining an appropriate BESS size for multiple usages is a critical planning problem for grid-interactive data centers.

This paper addresses the BESS sizing problem for grid-interactive data centers by developing an optimization-based framework that explicitly captures the multiple service requirements imposed on the storage system. The framework considers demand response support, smoothing of fast load fluctuations, reactive power capacity for quasi-static voltage support during point of common coupling (PCC) voltage sag events, and slower operational services such as net load stabilization and load ramping limitation. By representing these requirements within a unified planning model, the proposed approach determines the minimum BESS power and energy capacity needed to ensure that the data center can satisfy both internal operational constraints and external grid-support expectations.

The main contributions of this work are as follows:
\begin{itemize}
    \item We formulate a multi-capability BESS sizing problem for data centers that jointly considers demand response, fast load fluctuation smoothing, voltage support at PCC, and load ramping constraints.
    \item We distinguish between capabilities that consume energy and capabilities that primarily require power headroom, enabling a more realistic representation of BESS capability requirements.
    \item We develop a scenario-based optimization framework that sizes the BESS to satisfy different operating conditions, including demand response events and main-grid fault conditions.
\end{itemize}

The rest of this paper is organized as follows. Section~\ref{sec_problem_definition} describes the problem and the required grid-support functionalities. Section~\ref{sec_model} presents the mathematical formulation of the proposed BESS sizing model. Section~\ref{sec_case_study} discusses case studies and numerical results. Section~\ref{sec_conclusion} concludes the paper and outlines future research directions.

\section{Problem Description}
\label{sec_problem_definition}
Our goal is to determine the optimal size of the BESS to equip data centers with multiple capabilities, such as demand response, fast fluctuation smoothing, and voltage support during main grid fault, while considering operational requirements and constraints. These functionalities are primarily enabled by the paired BESS, with its rated power and energy capacity strategically allocated across different functions. Additionally, the data center manages a load queue for delay-insensitive tasks, offering further operational flexibility.

The demand response capacity is defined by the flexibility of the data center to decrease its load over a specified duration. This flexibility is expressed as a percentage of the data center's total rated load. Achieving this capability relies on both the rated power and energy capacity of the BESS.

Fast load fluctuations in data centers refer to sudden, rapid changes in power demand caused by variations in workloads. These fluctuations typically have high frequencies ($>$ 1 Hz) and large magnitudes, which could be up to 80\% of the total rated IT load. The BESS smooths these fluctuations by charging during low-load periods and discharging during high-load periods. In this study, we assume that these charging and discharging actions are balanced and occur on a subsecond timescale, yielding no net change in the BESS state of charge (SoC). If the bottom and top 10\% of SoC are reserved (i.e., not used during normal operation or for other services), the BESS can accommodate the energy exchange required for fast fluctuation smoothing. However, the required rated power to manage these fluctuations is significant, approximately 40\%–50\% of the data center's total rated load.

Voltage support capacity represents the reactive power capability required to maintain the calculated data center root node voltage within specified limits during an assumed PCC voltage sag. When the voltage at the PCC drops, the BESS provides reactive power to improve the root node voltage, helping preserve normal operation and protect connected equipment. Because this service is primarily reactive, it does not require net energy charging or discharging; however, it does require sufficient inverter apparent power capacity to supply the needed reactive power. Accordingly, the analysis establishes the BESS apparent power headroom required for quasi-static voltage support.

During daily operations, the BESS can stabilize the data center’s load over a timescale of minutes to hours and limit load ramping to stay within the required thresholds. To achieve these functionalities, the BESS must have sufficient rated power and energy capacity while ensuring proper state of charge (SoC) is maintained throughout its operation.

The detailed structure of the data center and its distribution system is not explicitly modeled. Instead, the focus is on aggregated load and BESS operation at minute timescale. The control of devices and system transient dynamics for fast fluctuation smoothing or VRT is not modeled or simulated. However, the model ensures that sufficient BESS rated power, energy capacity, and SoC are available to meet the requirements of these functionalities.

\section{Model Formulation}
\label{sec_model}
The model adopts a scenario-based framework to represent possible operation conditions. In each scenario, one or more DR events and PCC voltage sag events may occur, either individually or in combination. By constructing multiple scenarios, the model captures the possible variation in the timing and duration of those events. This setting enables the result to have sufficient large BESS under a wide range of plausible operating conditions, reflecting the uncertainty and diversity of real-world event occurrences.

\subsection{BESS size limits} 
We use decision variables $E^\mathrm{BESS}$ and $P^\mathrm{BESS}$ to denote the BESS energy capacity and rated power, respectively. Each of them has an upper limit as specified by the following constraints.
\begin{align}
& 0\leq E^\mathrm{BESS}\leq \overline E^\mathrm{BESS}\\
& 0\leq P^\mathrm{BESS}\leq \overline P^\mathrm{BESS}
\end{align}

The following operating constraints and decision variables are replicated across all scenarios $s\in\mathcal S$, as each scenario represents a distinct operating condition. To avoid repetition, we omit explicit index ranges in what follows: constraints having decision variables with $t$ are enforced for all time periods $t\in\mathcal T$ (within the optimization horizon and proper boundaries) and those having decision variables with $s$ are for all scenarios $s\in\mathcal S$.

\subsection{DC load operation constraints}
By strategically operating the BESS and managing a load queue, the DC guarantees IT workload is processed timely and its power consumption satisfies grid requirements.
\begin{align}
&p^\mathrm{q}_{t,s}=p^\mathrm{q}_{t-1,s}+d^\mathrm{IT}_t-p^\mathrm{IT}_{t,s}\label{load_queue}\\
&p^\mathrm{IT}_{t,s}\geq \hat d^\mathrm{IT}_t\label{load_sense}\\
&p^\mathrm{IT}_{t,s}+p^\mathrm{Cool}_{t,s}+p_{t}^\mathrm{ch_{op}}-p_{t,s}^\mathrm{dis_{op}}=p_{t,s}^\mathrm{DC}\label{load_bal}\\
&-P^\mathrm{ramp}\leq p_{t,s}^\mathrm{DC}-p_{t-1,s}^\mathrm{DC}\leq P^\mathrm{ramp}\label{load_ramp}\\
&p_{t,s}^\mathrm{DC}\leq \overline P^\mathrm{DC}-\alpha P^\mathrm{R}D^\mathrm{dr}_{t,s}\label{load_lim}\\
&p_{t,s}^\mathrm{DC}\leq (1+\delta)\sum_{\tau=t-T^\mathrm s}^{t+T^\mathrm s}p_{\tau,s}^\mathrm{DC}/(2T^\mathrm s+1)\label{ma_smooth_up}\\
&p_{t,s}^\mathrm{DC}\geq (1-\delta)\sum_{\tau=t-T^\mathrm s}^{t+T^\mathrm s}p_{\tau,s}^\mathrm{DC}/(2T^\mathrm s+1)\label{ma_smooth_low}\\
&0\leq p^\mathrm{q}_{t,s}\leq \overline P^\mathrm{Q}\label{lim_queue}
\end{align}
Parameters $d^\mathrm{IT}_t$ and $\hat d^\mathrm{IT}_t$ represent total and time sensitive load at time $t$, respectively; $P^\mathrm{ramp}$ denotes the load ramping limit; $\overline P^\mathrm{DC}$ specify the upper limits of the DC; $P^\mathrm{R}$ denote the DC's rated load; $D^\mathrm{dr}_{t,s}$ is a 0-1 time series for demand response request signals in scenarios $s$; $\delta$ specifies the allowable relative load deviation from the local average; $2T^\mathrm s$ is half-width of the smoothing window in number of time steps; $\overline P^\mathrm{Q}$ denotes the upper limit of unprocessed load in the queue. Decision variables $p^\mathrm{q}_{t,s}$, $p^\mathrm{IT}_{t,s}$, and $p_{t,s}^\mathrm{DC}$ represent the unprocessed load in the queue, processed load, and total load of the DC, respectively; $p^\mathrm{Cool}_{t,s}$ denotes the load of the cooling system; the BESS charged and discharged power during DC operation are represented by $p_{t,s}^\mathrm{ch_{op}}$ and $p_{t,s}^\mathrm{dis_{op}}$, respectively.

Constraint (\ref{load_queue}) is the dynamic equation of the load queue; constraint (\ref{load_sense}) guarantees time sensitive load is processed timely; constraint (\ref{load_bal}) is load balance equation; constraint (\ref{load_ramp}) is DC load ramping limit; constraint (\ref{load_lim}) limits DC maximum load with and without demand response request; constraints (\ref{ma_smooth_up}-\ref{ma_smooth_low}) reduce short-term fluctuation by a moving average method; constraint (\ref{lim_queue}) imposes the limits for unprocessed load in the queue.

\subsection{BESS operation for demand response}
The DC has the flexibility to adjust $\alpha$ of its rated load level $P^\mathrm{R}$ for $T^\mathrm{dr}$ hours by operating the BESS. The following constraints model this demand response capability 
\begin{align}
&p^\mathrm{ch_{dr}}=\alpha P^\mathrm{R}\label{DR_ch}\\
&p^\mathrm{dis_{dr}}=\alpha P^\mathrm{R}\label{DR_dis}\\
&p_{t,s}^\mathrm{ch_{op}}-p_{t,s}^\mathrm{dis_{op}}+p^\mathrm{ch_{dr}}\leq P^\mathrm{Batt}\label{ch_lim}\\
&p_{t,s}^\mathrm{ch_{op}}-p_{t,s}^\mathrm{dis_{op}}-p^\mathrm{dis_{dr}}\geq -P^\mathrm{Batt}\label{dis_lim}\\
&E^\mathrm{dr}=\alpha P^\mathrm{R}T^\mathrm{dr}\label{DR_E}
\end{align}
The decision variables $p_{t,s}^\mathrm{ch_{dr}}$ and $p_{t,s}^\mathrm{dis_{dr}}$ represent the BESS charging and discharging power allocated for providing load flexibility, respectively; $E^\mathrm{dr}$ denotes the energy required to support such demand response over a duration of $T^\mathrm{dr}$ hours.

Constraints (\ref{DR_ch}-\ref{DR_dis}) determine the rated power required for demand response; constraints (\ref{ch_lim}-\ref{dis_lim}) impose limits on the BESS rated power for DC demand response and operation; constraint (\ref{DR_E}) calculates the energy required to support demand response.

\subsection{BESS for fast load fluctuation smoothing}
To support fast load fluctuation smoothing, the BESS requires a rated power, $P^\mathrm{fs}$, calculated as a fraction, $\beta$, of the DC's rated load level. This relationship is represented by the following constraint. The parameter $\beta$ is tunable and can be adjusted based on the fluctuation characteristics of the DC's workloads.
\begin{align}
P^\mathrm{fs}=\beta P^\mathrm{R}
\end{align}

\subsection{BESS state dynamic constraints}
BESS operation and dynamic constraints are as following.
\begin{align}
&SoC_{t,s}-SoC_{t-1,s}=(\eta^\mathrm{ch}p_{t,s}^\mathrm{ch_{op}}-\eta^\mathrm{dis}p_{t,s}^\mathrm{dis_{op}})*T^\mathrm{L}\label{batt_dyn}\\
&0\leq p_{t,s}^\mathrm{ch_{op}}\leq P^\mathrm{Batt}z^\mathrm{op}\label{op_ch_lim}\\
&0\leq p_{t,s}^\mathrm{dis_{op}}\leq P^\mathrm{Batt}(1-z^\mathrm{op})\label{op_dis_lim}\\
&0.1E^\mathrm{BESS}+E^\mathrm{dr}(1-D^\mathrm{dr}_{t,s})\leq SoC_{t,s}\leq 0.9E^\mathrm{BESS}\label{soc_lim}
\end{align}
Parameters $\eta^\mathrm{ch}$ and $\eta^\mathrm{dis}$ represent BESS charge and discharge efficiency. Variable $SoC_{t,s}$ denotes BESS SoC.

Constraint (\ref{batt_dyn}) is BESS SoC dynamic equation; constraints (\ref{op_ch_lim}-\ref{op_dis_lim}) guarantee BESS will not charged and discharged simultaneously; constraint (\ref{soc_lim}) is BESS SoC limits considering DC operation and demand response. The SoC is restricted to 10\% to 90\% of its total energy capacity to avoid being over charged or discharged, which also allows slightly fluctuation of SoC due to fast load fluctuation smoothing. 

Constraint (\ref{batt_dyn}) represents the dynamic equation for the BESS SoC; constraints (\ref{ch_lim}-\ref{dis_lim}) ensure that the BESS does not charge and discharge simultaneously; constraint (\ref{soc_lim}) enforces SoC limits while accounting for DC operation and demand response. The SoC is restricted to remain between 10\% and 90\% of the total energy capacity to prevent overcharging or overdischarging. This range also accommodates minor SoC fluctuations caused by fast load fluctuation smoothing.

\subsection{DC PCC voltage equations}
The constraints for voltage support capability are formulated below. We aggregate the DC as a single point and consider then transmission line from PCC to the root node of DC distribution system.
\begin{align}
&p^l_{t,s}=p_{t,s}^\mathrm{DC}\label{bal_real}\\
&q^l_{t,s}=q_{t,s}^\mathrm{vs}\label{bal_reac}\\
&V^0_t-\gamma F^\mathrm{vs}_{t,s}-V^\mathrm{DC}_{t,s}=rp^l_{t,s}+xq^l_{t,s}\label{power_flow}\\
&\left(P^\mathrm{Batt}+P^\mathrm{fs}\right)^2+\left(q_{t,s}^\mathrm{vs}\right)^2\leq\left(P^\mathrm{BESS}\right)^2\label{bess_power}\\
&V^\mathrm{ref}-\Delta V\leq V^\mathrm{DC}_{t,s}\leq V^\mathrm{ref}+\Delta V\label{DC_vol}
\end{align}
Parameter $V^0_t$ denotes the normal voltage at the PCC; $V^\mathrm{ref}$ represents the reference voltage at the root node of the DC distribution grid; $\Delta V$ specifies the maximum allowable deviation from this reference voltage; $r$ and $x$ correspond to the resistance and reactance of the main transmission line connecting the PCC to the DC’s root node. $F^\mathrm{vs}_{t,s}$ is a 0-1 time series indicating if PCC voltage sag happens in scenario $s$. It it takes 1, the voltage at PCC drop by $\gamma$. The decision variables $p^l_{t,s}$ and $q^l_{t,s}$ denote the real and reactive power flowing through the main transmission line, respectively; $q_{t,s}^\mathrm{vs}$ represents the reactive power supplied by the BESS; $V^\mathrm{DC}_{t,s}$ denotes the voltage at the root node of the DC distribution grid.

Constraints (\ref{bal_real}-\ref{bal_reac}) ensure the balance of real and reactive power at the root node of the DC distribution grid, respectively; constraint (\ref{power_flow}) defines the power flow equation for the main transmission line; constraint (\ref{bess_power}) imposes the total power limit of the BESS; constraint (\ref{DC_vol}) sets the voltage limit at the root node of the DC distribution grid.

\subsection{DC cooling power equation}
Let PUE be the DC's power usage effectiveness, and assume nonessential load is negligible. Under these assumptions, the cooling power can be approximated as follows:
\begin{align}
&p^\mathrm{Cool}_{t,s}=(\mathrm{PUE}-1)p^\mathrm{IT}_{t,s}
\end{align}

\subsection{Objective function}
In this study, we consider two situations. In the first one, the demand response capacity $\alpha$ and maximum allowed voltage drop at PCC $\gamma$ are given parameters. These can be viewed as grid requirements for the DC to get connected. The objective in this situation is to minimize the total cost of the BESS to satisfy these requirements. In the second one, both $\alpha$ and $\gamma$ are treated as decision variables. The problem is formulated as a multi-objective optimization that minimizes BESS cost and maximizes demand response and voltage support capacities. The Pareto front will be calculated for the decision maker to balance the objectives.
\subsubsection{Single objective formulation}
Let $C^E$ and $C^P$ represent the unit costs of energy capacity and rated power for the BESS, respectively. The objective for minimizing BESS cost is as below.
\begin{align}
\min\quad & C^EE^\mathrm{BESS}+C^PP^\mathrm{BESS}\nonumber\\
&+\sum_{t\in\mathcal T}\sum_{s\in\mathcal S}|V^\mathrm{DC}_{t,s}-V^\mathrm{ref}|\label{single_obj}
\end{align}
The first two terms are the cost for BESS energy capacity and rated power, respectively. The last one is a penalty term for voltage deviation from the reference value.

The BESS size pair $(E^\mathrm{BESS},P^\mathrm{BESS})$ will be sufficiently large for all scenarios, while the BESS is operated differently according to their respective demand response request and main grid fault signals. 

\subsubsection{Multi-objective formulation}
A common method to solve multi-objective optimization is the weighted sum approach. Let $w^\mathrm{BESS}$, $w^\mathrm{DR}$, and $w^\mathrm {VS}$ be the weight factors for BESS cost, demand response capacity, and voltage support capacity, respectively. The scalarized objective can be expressed as follows.
\begin{align}
\min\quad & w^\mathrm{BESS}(C^EE^\mathrm{BESS}+C^PP^\mathrm{BESS})-w^\mathrm{DR}\alpha\nonumber\\
&-w^\mathrm {VS}\gamma+\sum_{t\in\mathcal T}\sum_{s\in\mathcal S}|V^\mathrm{DC}_{t,s}-V^\mathrm{ref}|\label{multi_obj_weighted}
\end{align}
By repeated solving the resulting problem with different combinations of weight factor values, the Pareto front can be derived. DC designers can select the desired solution from the Pareto front that balances BESS cost and DR and voltage support capacities for their needs.

\section{Case Study}
\label{sec_case_study}
\subsection{Usecase setting}
\begin{figure}
    \centering
    \includegraphics[width=1\linewidth]{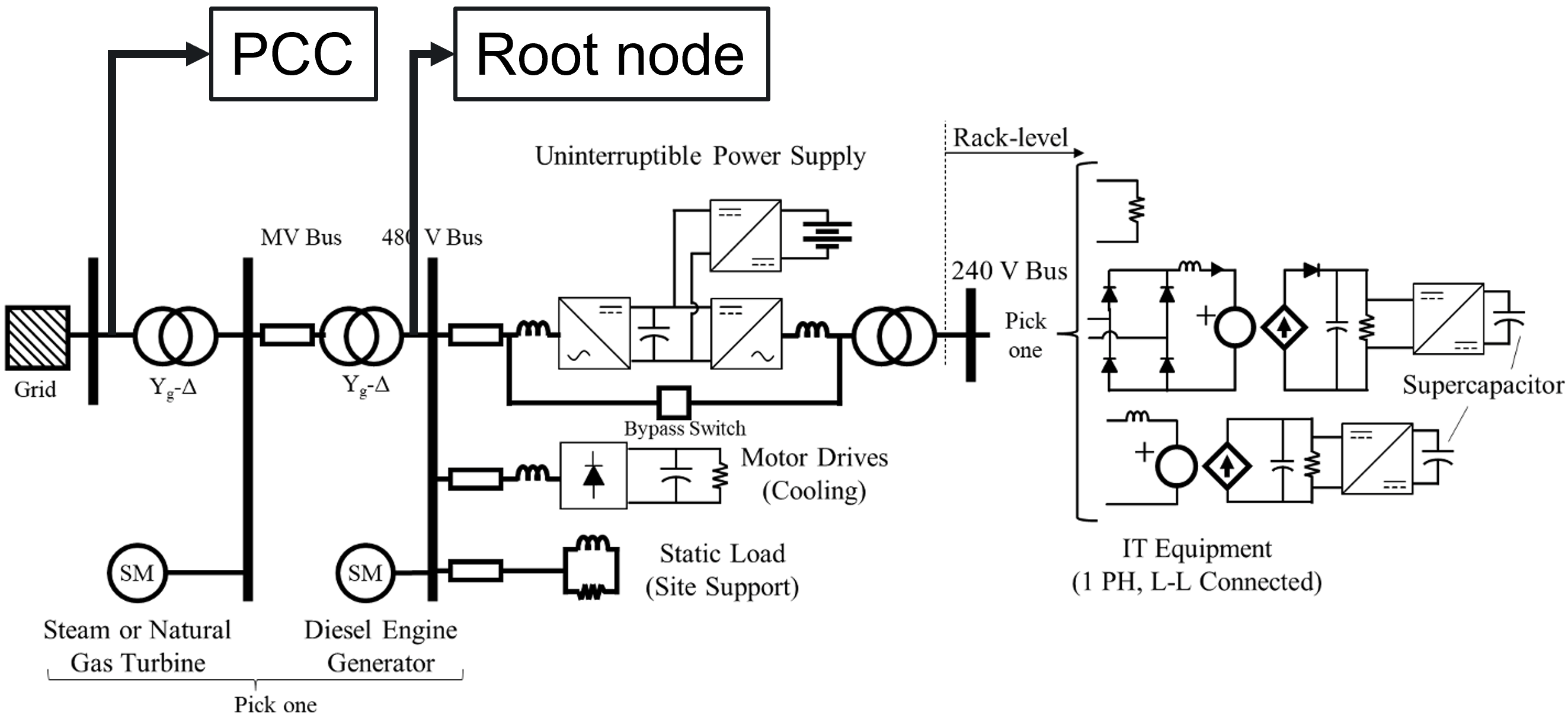}
    \caption{Data center grid connection and power supply architecture. Root node connects to the main grid through the PCC. Total DC load and the aggregate charging/discharging power of all BESS units are represented at the root node.}
    \label{fig:DC_structure}
\end{figure}

The structure of the DC system is illustrated in Fig.~\ref{fig:DC_structure}. The root node is connected to the main grid through the PCC. The total DC load and the aggregate charging/discharging power of all BESS units are represented at the root node. The calculated BESS size may represent the combined capacity of multiple devices distributed throughout the DC system; however, the network details beyond the root node (i.e., the internal DC system topology) are not modeled in this paper.

\begin{figure}
    \centering
    \includegraphics[width=0.8\linewidth]{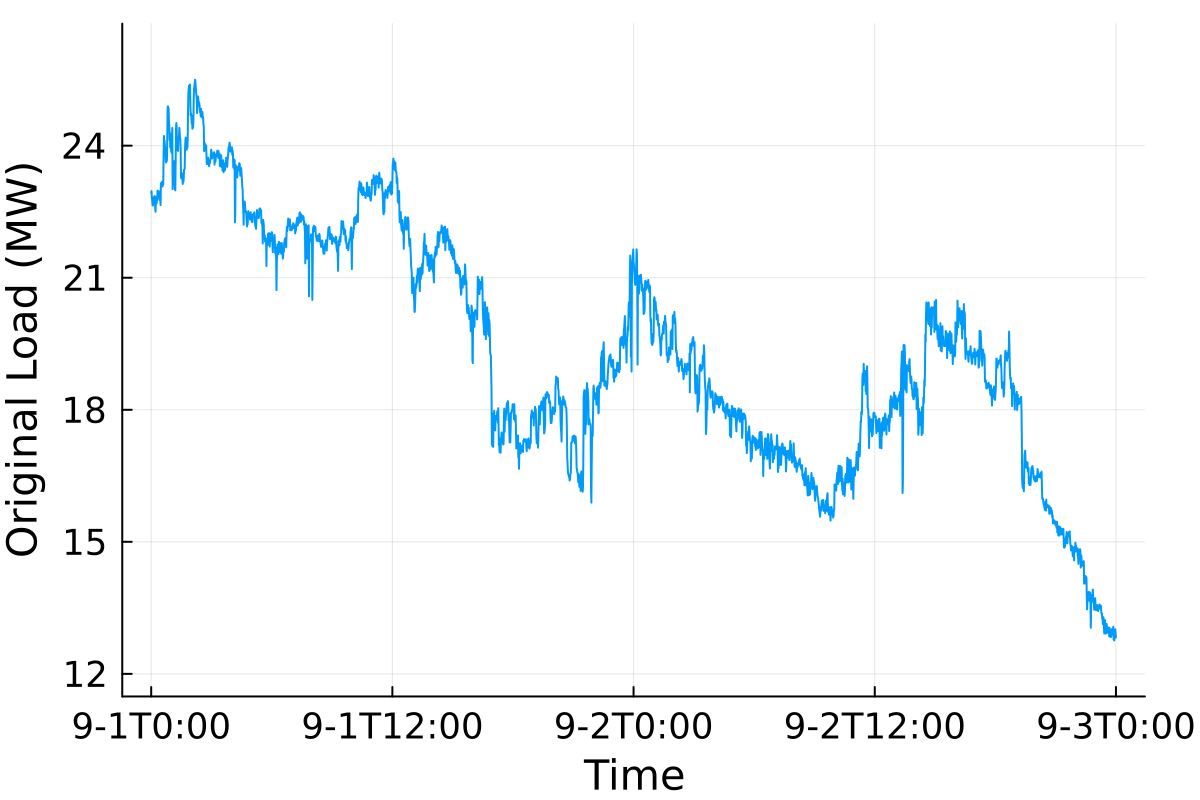}
    \caption{Data center IT load profile over a 48-hour period at 1-minute resolution.}
    \label{fig:load}
\end{figure}

The original DC load data used in this case study are taken from the MIT Supercloud dataset~\cite{samsi2021supercloud}. The total IT load over a 48-hour period, from Sep.~1 at 00:00 to Sep.~2 at 23:59, is retrieved, resampled to 1-minute resolution, and rescaled. The resulting load profile is shown in Fig.~\ref{fig:load}. We assume that 80\% of the load is time-sensitive and therefore cannot be deferred. The PUE is set to 1.3 and rated load is set to 35 MW. The upper limit for total load is 30 MW.

The cost of BESS is estimated using PNNL's Energy Storage Evaluation Tool~\cite{pnnl_eset}. Specifically, the unit energy capacity cost $C^E=\$370/\mathrm{kWh}$ and the unit rated power cost $C^P=\$105/\mathrm{kW}$.

In most of the cases, two representative scenarios are considered. In the DR scenario, two demand response events are scheduled at 7:00 and 22:00 on Sep.~1. During each event, the DC reduces its load for one hour. After each event, a six-hour interval with no additional demand response is assumed, allowing the BESS to recharge. In the PCC voltage sag scenario, voltage drops occur at 3:20 and 16:40 on Sep.~1 and last for 20 minutes. Maximum allowed voltage deviation $\Delta V$ at the root node is set to 0.2~p.u. These event timings and durations are simplified assumptions for the case study. While some performance and sensitivity analysis will be conducted, more realistic demand response schedules and voltage sag durations would require detailed transmission system simulations, therefore is out the scope of this paper.

The following four cases are analyzed. The resulting BESS sizing solutions and associated costs are summarized in Table~\ref{tab:single_obj_sol}.
\begin{itemize}
\item \textbf{Demand response only}: Only the DR scenario is considered, and the single objective function (\ref{single_obj}) is optimized. When requested, the DC reduces its peak load by 20\% of its rated load (i.e., $\alpha=0.2$).
\item \textbf{Voltage sag (VS) only}: Only the voltage sag scenario is considered, and the single objective function (\ref{single_obj}) is optimized. The PCC voltage is assumed to drop to 0.6~p.u.
\item \textbf{Integrated DR and VS}: Both DR and voltage sag scenarios are considered, using the same parameter values as in the corresponding single scenario cases. The performance of resulting BESS size is analyzed by changing the interval of DR events in the DR scenario. In addition, BESS size sensitivity is tested with more randomly generated scenarios.
\item \textbf{Multi-objective}: Both demand response and voltage sag scenarios are considered, and the scalarized objective (\ref{multi_obj_weighted}) is optimized under different weight factor combinations. The demand response capacity $\alpha$ is constrained to $[0.1,\,0.3]$, and allowed PCC voltage drop $\gamma$ is constrained to $[0.2~\text{p.u.},\,0.4~\text{p.u.}]$.
\end{itemize}

\begin{table}[htbp]
  \centering
  \caption{BESS Size Solutions and Cost for Single Objective Cases}
  \resizebox{\linewidth}{!}{
    \begin{tabular}{lccc}
    \toprule
          & \multicolumn{1}{l}{$E^\mathrm{BESS}$ (MWh)} & \multicolumn{1}{l}{$P^\mathrm{BESS}$ (MW)} & \multicolumn{1}{l}{Cost (\$M)} \\
    \midrule
    DR Only & 18.69 & 30.28 & 10.09 \\
    VS Only & 9.94  & 78.54 & 11.93 \\
    DR + VS & 18.69 & 80.82 & 15.40 \\
    \bottomrule
    \end{tabular}}%
  \label{tab:single_obj_sol}%
\end{table}%

\subsection{Demand response scenario only}
As shown in the first row of Table~\ref{tab:single_obj_sol}, when only the demand response scenario is considered, the optimal BESS size is 30.28~MW/18.69~MWh, with a total cost of \$10.09~million. Comparing to the voltage support only case, providing demand response requires larger energy capacity. Since main grid fault does not happen in this case, the BESS is not used for providing reactive power to support root node voltage, resulting in smaller BESS power rating.

\begin{figure}
    \centering
    \includegraphics[width=1\linewidth]{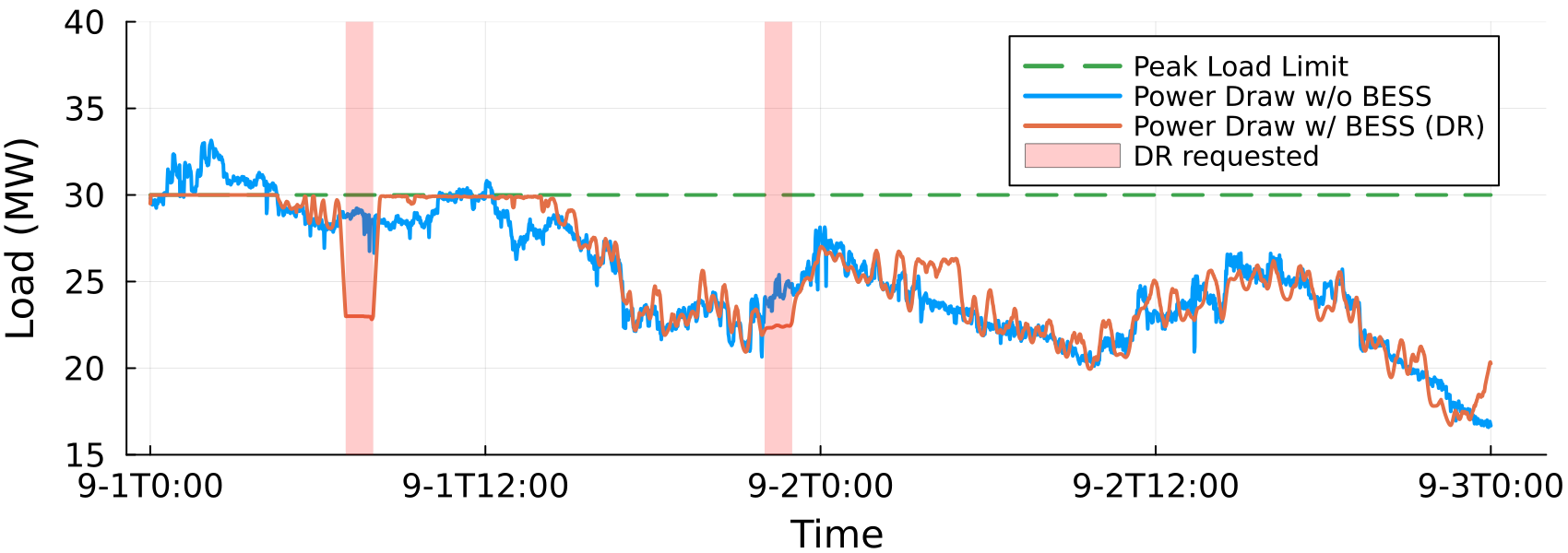}
    \caption{Data center net load in demand response only case. The hours that demand response events are requested are marked by pink color.}
    \label{fig:load_DR}
\end{figure}

\begin{figure}
    \centering
    \includegraphics[width=1\linewidth]{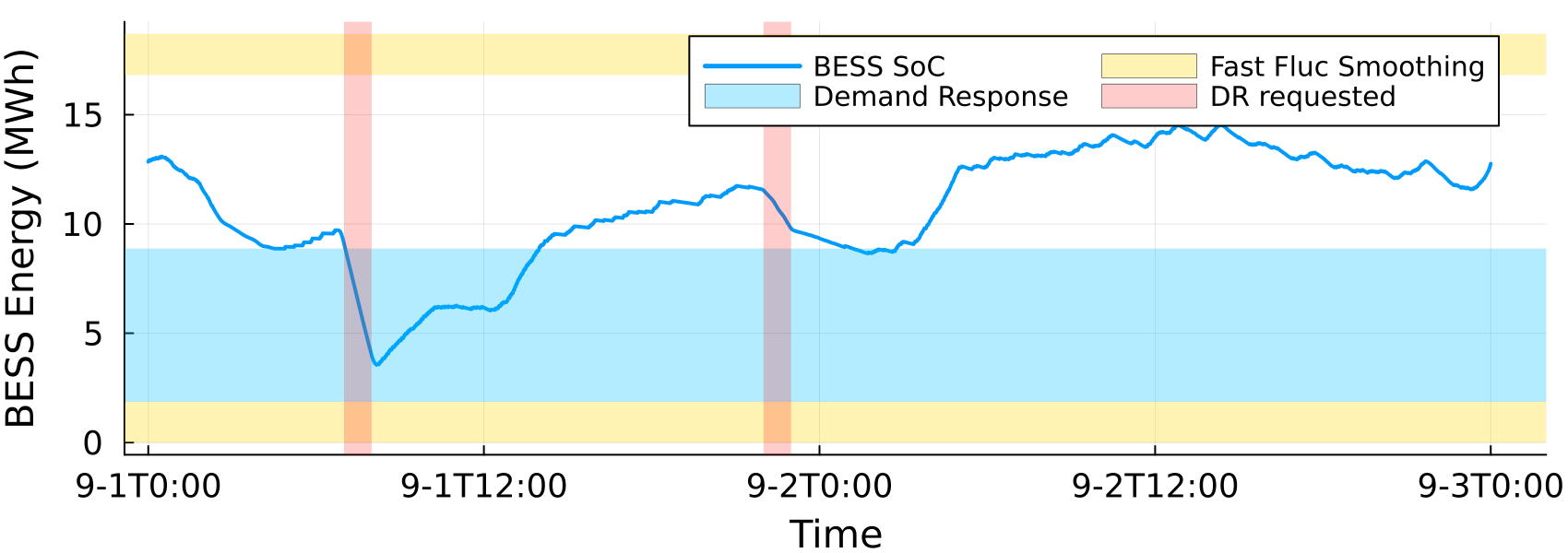}
    \caption{BESS energy capacity allocation and SoC trajectory in demand response only case. The energy capacity allocated for demand response is marked by blue. The yellow parts represent the upper and lower 10\% of total BESS energy capacity for accommodating energy exchange for fast fluctuation smoothing.}
    \label{fig:SoC_DR}
\end{figure}

Fig.~\ref{fig:load_DR} illustrates DC's initial total load and power draw with BESS, and Fig.~\ref{fig:SoC_DR} shows the BESS energy capacity allocation and SoC trajectory. It can be observed in Fig.~\ref{fig:load_DR} that with the BESS, DC's total load is restricted below the upper limit of 30 MW, and is able to further reduce to 23 MW when demand response is requested. The load curve with BESS is smoother than that without BESS. Based on constraint (\ref{DR_E}), 7 MWh is reserved in the BESS to reduce peak load during demand response events, which is the blue part in Fig. ~\ref{fig:SoC_DR}. At the beginning of Sep.~1, DC's load is high and the BESS is discharged to meet the 30 MW load limit, but the reserved energy for demand response is not used due to the SoC limit constraint (\ref{soc_lim}). This ensures that enough energy in BESS is available whenever demand response is requested. The yellow parts in Fig.~\ref{fig:SoC_DR} represent the upper and lower 10\% of total BESS energy capacity, which is not used for regular operation for safety and accommodates energy exchange for fast fluctuation smoothing.

\subsection{Voltage sag scenario only}
When only the voltage sag scenario is considered, the optimal BESS size is 78.54~MW/9.94~MWh, with a total cost of \$11.93~million, as shown in the second row of Table~\ref{tab:single_obj_sol}. Because demand response is not included, the required energy capacity is much smaller than in the demand response only case. However, the required power rating is substantially larger to provide voltage support during voltage sags.

\begin{figure}
    \centering
    \includegraphics[width=1\linewidth]{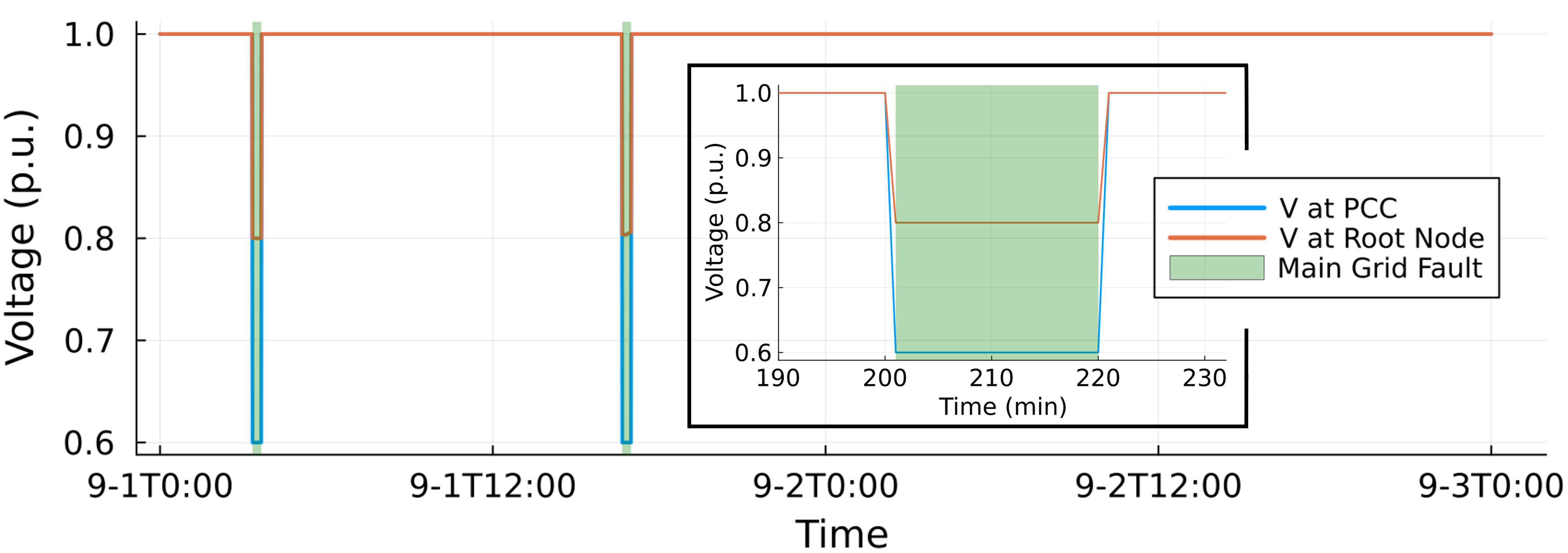}
    \caption{PCC and data center root node voltage in the voltage sag only case. The interval from Sep.~1 3:10~am to 3:50~am is zoomed in.}
    \label{fig:V_VFRT}
\end{figure}

Fig.~\ref{fig:V_VFRT} shows the voltages at the PCC and the DC root node; the interval from Sep.~1 3:10~am to 3:50~am is zoomed in to highlight the fault response. During the event, the PCC voltage drops to 0.6~p.u., whereas the root-node voltage remains above 0.8~p.u., satisfying the maximum allowable deviation of 0.2~p.u.

\begin{figure}
    \centering
    \includegraphics[width=1\linewidth]{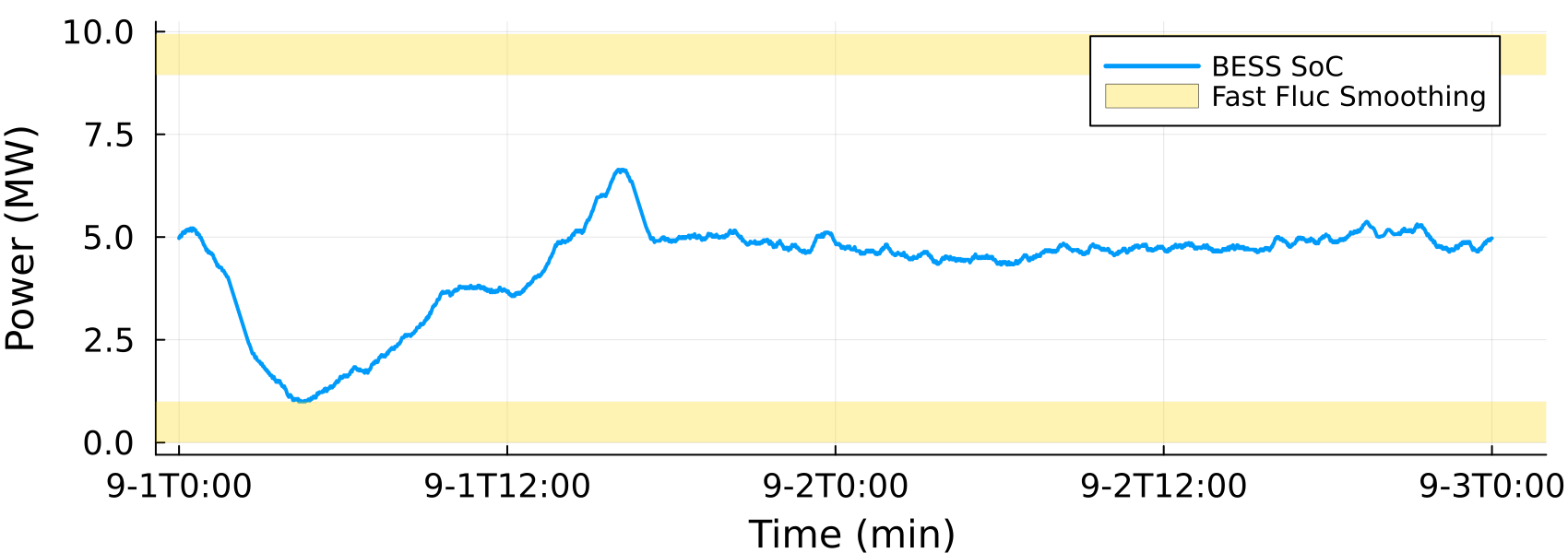}
    \caption{BESS SoC trajectory in voltage sag only case. The yellow parts represent the upper and lower 10\% of total BESS energy capacity for accommodating energy exchange for fast fluctuation smoothing.}
    \label{fig:SoC_VFRT}
\end{figure}

Fig.~\ref{fig:SoC_VFRT} illustrates the BESS SoC trajectory. Initially, the BESS discharges to enforce the 30~MW upper limit on the total load. Thereafter, the SoC remains close to 50\% for the remainder of the simulation horizon. Since no demand response events are considered, the BESS experiences no further substantial charging or discharging in this case.

\subsection{Integrating demand response and voltage sag scenarios}
When both demand response and voltage sag scenarios are included, the BESS provides both load flexibility and root node voltage support, which requires increases in both rated power and energy capacity. The resulting optimal BESS size is 80.82~MW/18.69~MWh, with a total cost of \$15.4~million. In this case, the net load profile, PCC and root-node voltage trajectories, and BESS SoC evolution are similar to those observed in the corresponding single-scenario cases. For brevity and to avoid redundancy, the figures are omitted.

The performance of the 80.82~MW/18.69~MWh BESS is evaluated under a set of extreme scenarios with shortened intervals between DR request events. Specifically, following the first DR event at 7:00, we assume that the second DR event is requested after 1, 2, \ldots, 5 hours. In the design problem, the BESS is assumed to have a 6-hour recharge period before the next DR request; therefore, a 20\% peak load reduction is guaranteed only for the first DR event. For the second DR event, the BESS reduces the peak load to the greatest extent possible given its remaining SoC and the limited time available for recharging.

\begin{figure}
    \centering
    \includegraphics[width=0.48\linewidth]{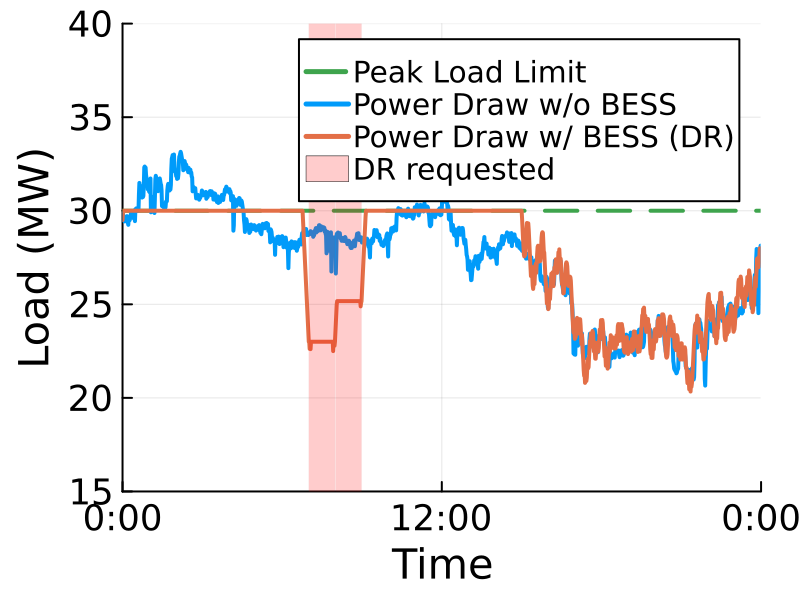}
    \includegraphics[width=0.48\linewidth]{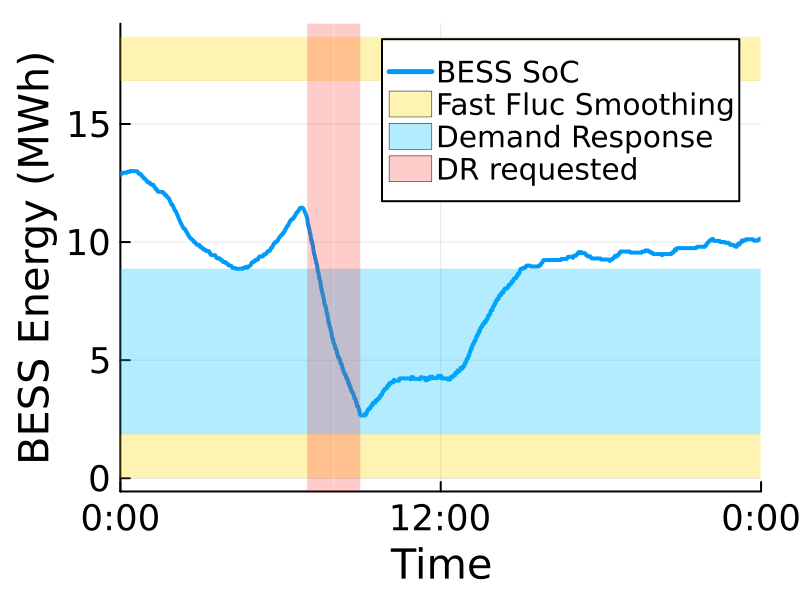}
    \includegraphics[width=0.48\linewidth]{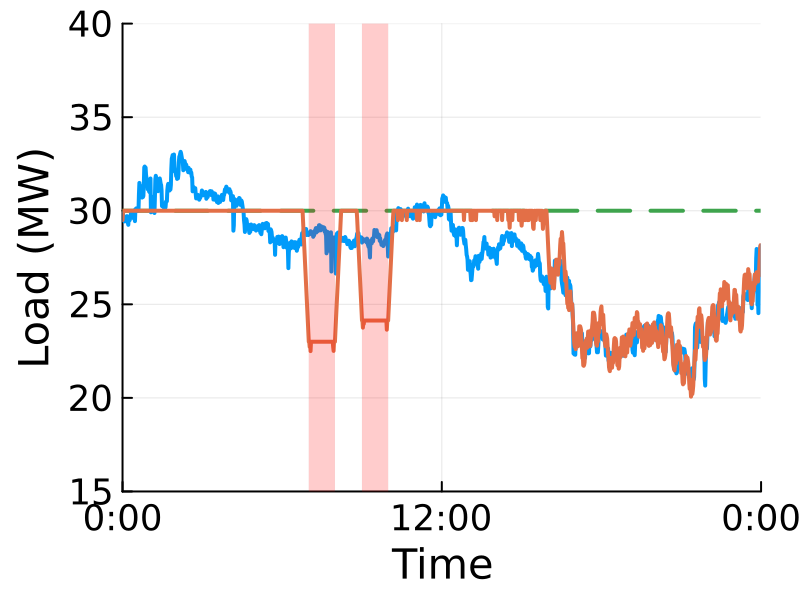}
    \includegraphics[width=0.48\linewidth]{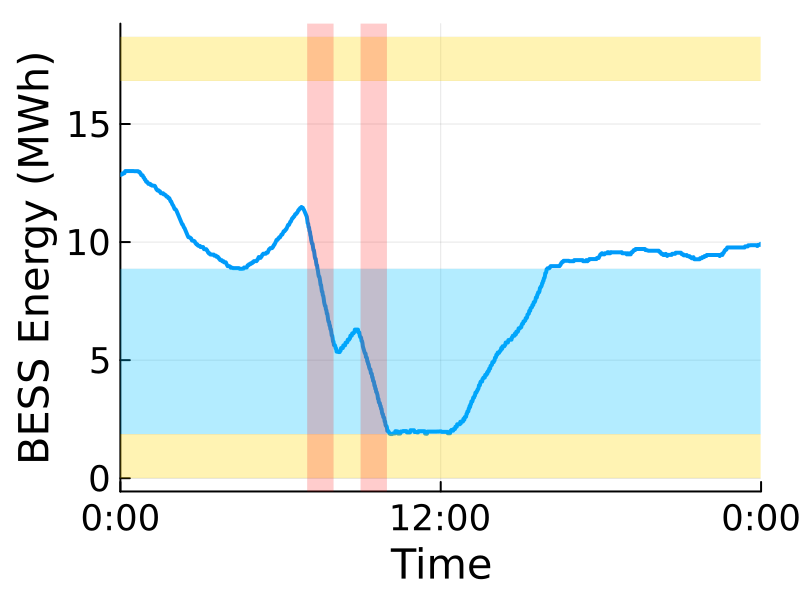}
    \includegraphics[width=0.48\linewidth]{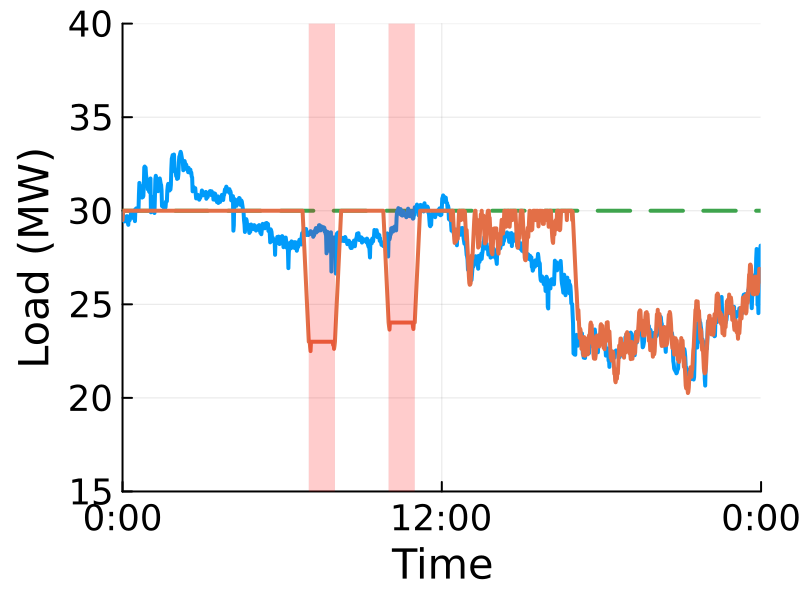}
    \includegraphics[width=0.48\linewidth]{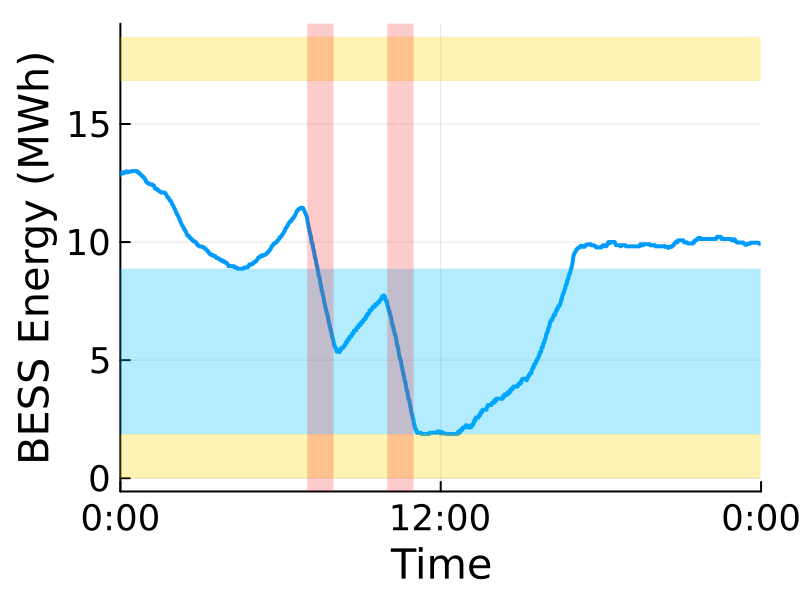}
    \includegraphics[width=0.48\linewidth]{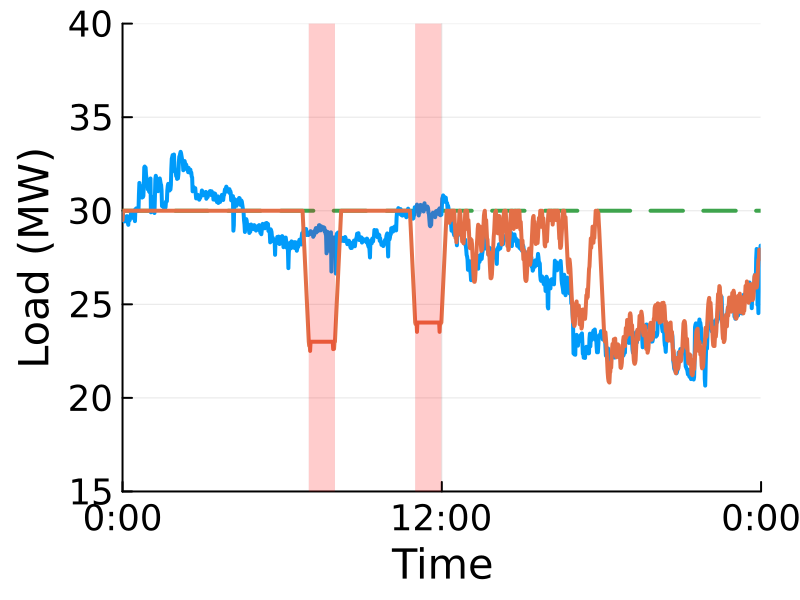}
    \includegraphics[width=0.48\linewidth]{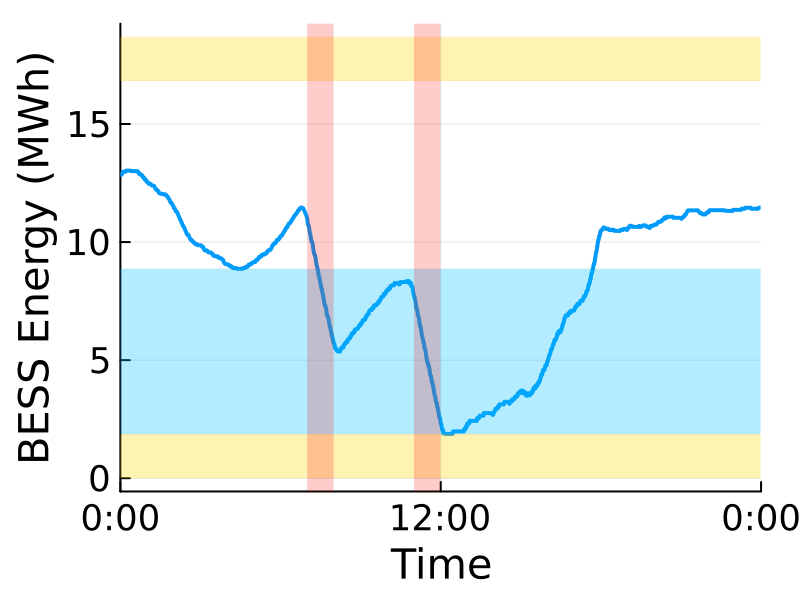}
    \includegraphics[width=0.48\linewidth]{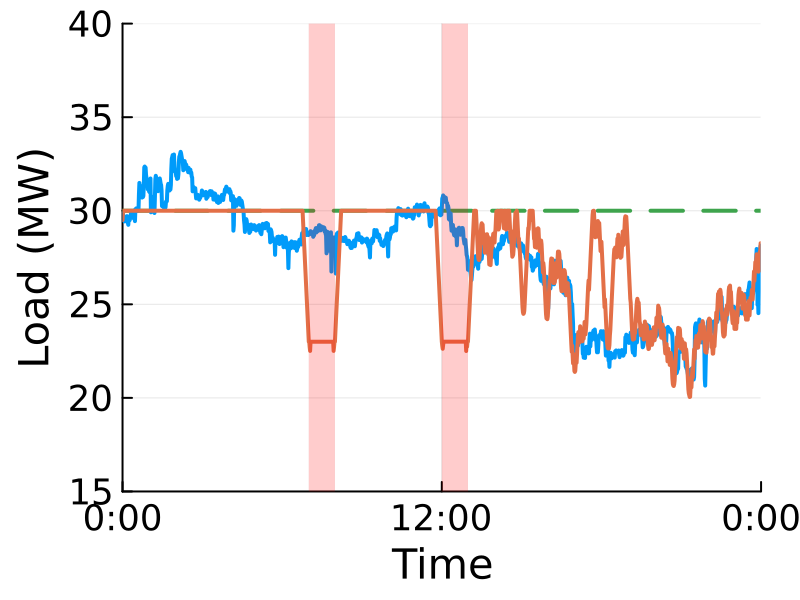}
    \includegraphics[width=0.48\linewidth]{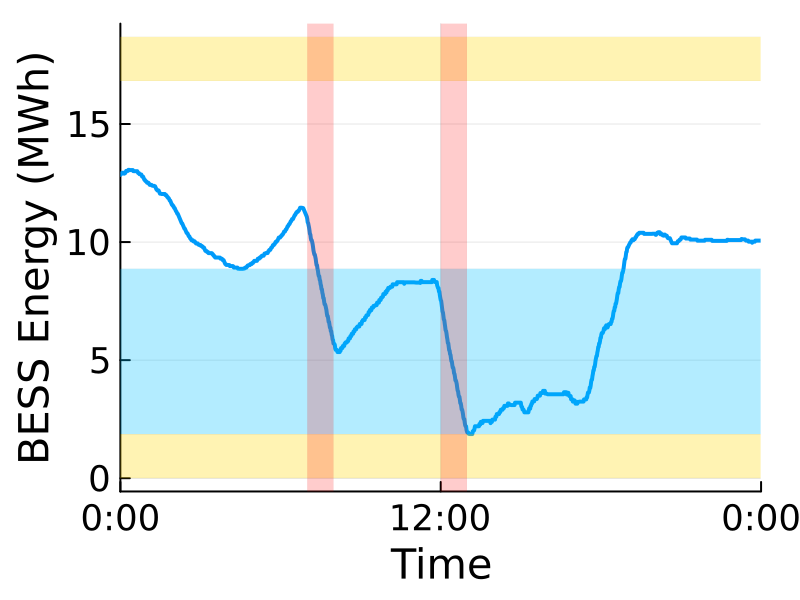}
    \includegraphics[width=0.48\linewidth]{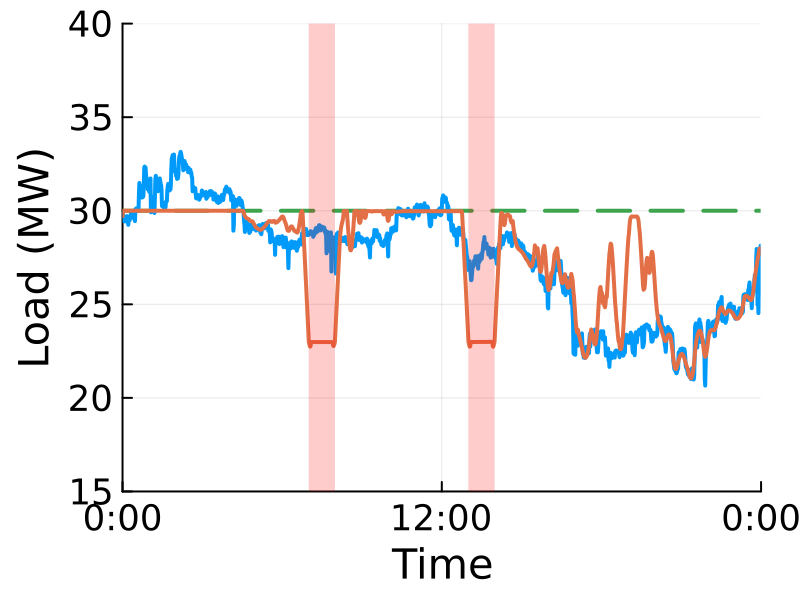}
    \includegraphics[width=0.48\linewidth]{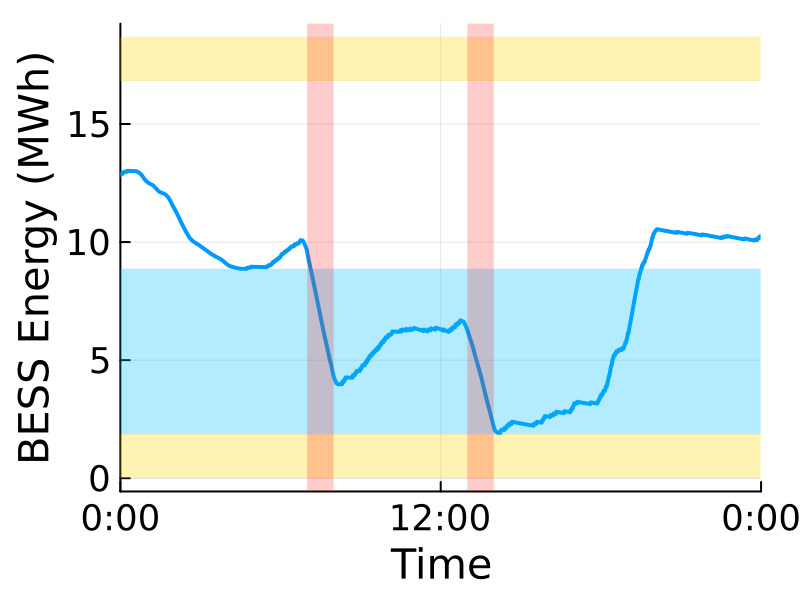}
    \caption{Data center's net load and BESS SoC trajectories under extreme demand response scenarios. In the second DR event, the peak load reduction is less than 20\% because the BESS has insufficient time to fully recharge.}
    \label{fig:DR_small_interval}
\end{figure}

Figure~\ref{fig:DR_small_interval} shows the DC's net load and BESS SoC trajectories. The results indicate that when the available recharge time is less than 3 hours, the peak load reduction achieved during the second DR event falls below 20\%. In all scenarios, the BESS fully depletes the energy reserved for DR after the second DR event is requested. Moreover, because the original DC's load reaches the maximum allowable value of 30~MW between approximately 10:00 and 12:00, the BESS cannot be recharged during this period, which further limits its ability to support the second DR event. These results suggest that guaranteeing load flexibility is highly challenging when DR events occur frequently during periods of high DC load.

We examine the sensitivity of the BESS size to DR and voltage sag events by solving the design problem under 15 randomly generated scenario configurations. Each configuration consists of two scenarios, and each scenario includes two events. For each configuration, four randomly generated events---two DR events and two VS events---are allocated across the two scenarios. All DR events are assumed to last 1 hour, whereas all VS events have a duration of 20 minutes.

\begin{figure*}
    \centering
    \includegraphics[width=0.9\linewidth]{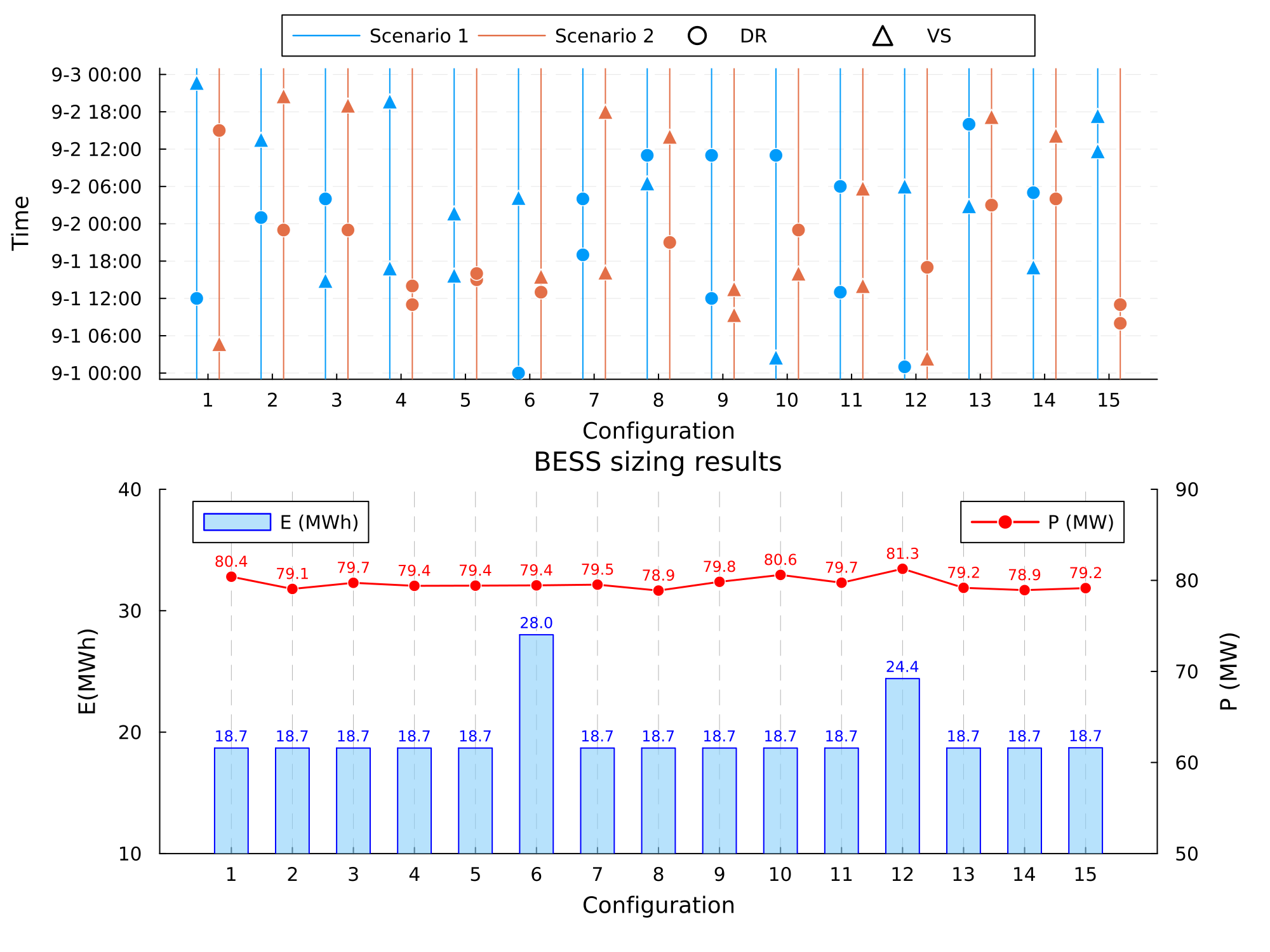}
    \caption{Randomly generated scenario configurations and corresponding BESS size in design results. Upper subplot illustrates the occurrence times and types of the two events in each scenario for all configurations. Lower subplot shows BESS rated power and energy capacity results for each configuration.}
    \label{fig:multi_sce_sol}
\end{figure*}

The random generated scenarios, including detailed information on the event types and their occurrence times in each configuration, and design results are illustrated in Fig~\ref{fig:multi_sce_sol}. For most configurations, the optimal BESS size is approximately 80~MW/18.7~MWh, indicating that a BESS of this size is sufficient to support the DC in most cases. The only exceptions are configurations 6 and 12. In these two cases, one DR event occurs at the beginning of Sep.~1, when the original load is high, requiring the BESS to discharge more energy to achieve peak load reduction and provide demand response support simultaneously.

\subsection{Multi-objective with both demand response and voltage sag scenarios}
\begin{figure}
    \centering
    \includegraphics[width=0.7\linewidth]{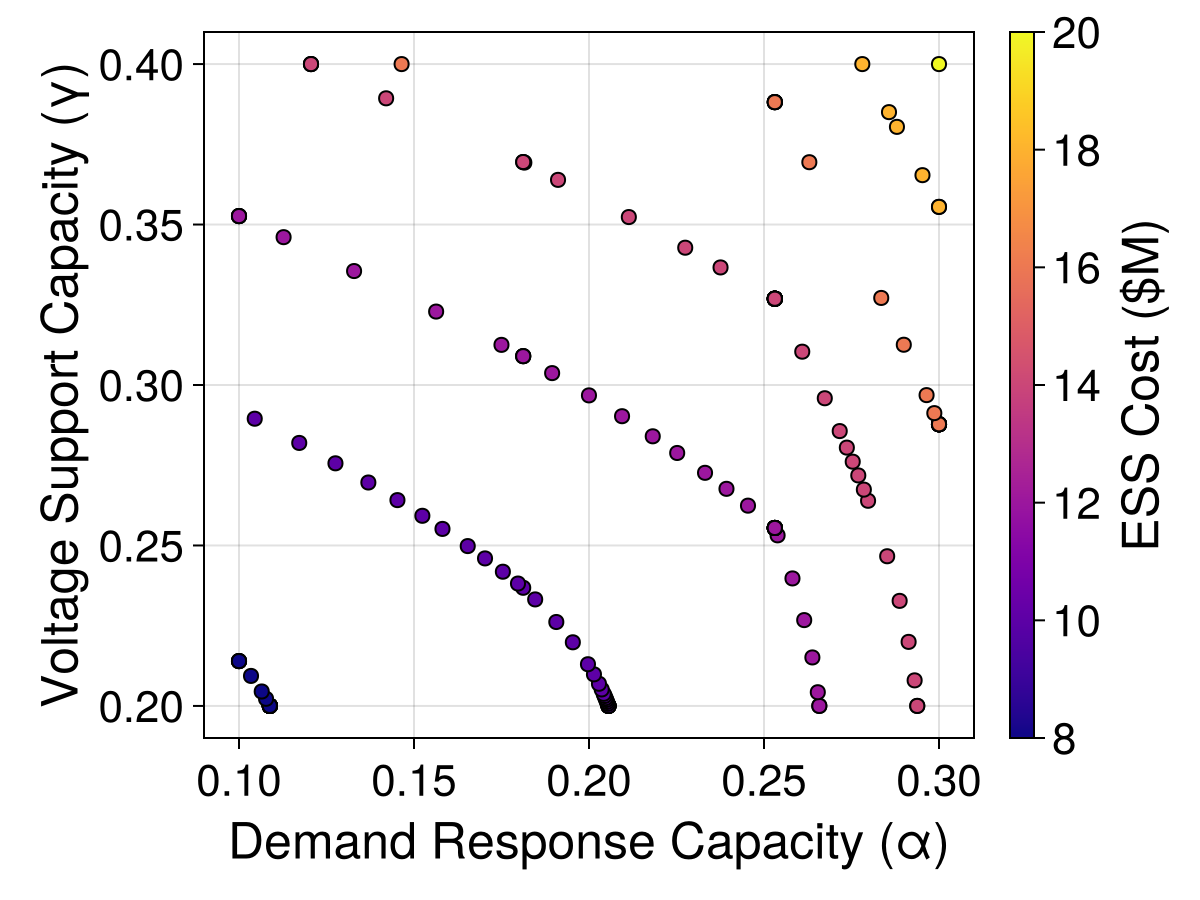}
    \caption{Pareto front of multiple objectives. The x-axis is demand response capacity, y-axis represents voltage support capacity, and the color of the dots indicates total BESS cost.}
    \label{fig:pareto_front}
\end{figure}

In the final case, the demand response capacity $\alpha$ and the allowable PCC voltage drop $\gamma$ are treated as decision variables. To make it easier to derive the Pareto front, an upper limit for total BESS cost in included. By varying BESS cost limit and the weight factors $w^\mathrm {DR}$ and $w^\mathrm {VS}$ in the scalarized objective function~(\ref{multi_obj_weighted}), the Pareto front is obtained, as shown in Fig.~\ref{fig:pareto_front}.

With a limited BESS budget, both the achievable demand response capacity and the voltage support capability are constrained. As the BESS cost increases, the DC can reduce more net load during demand response events and tolerate deeper PCC voltage sags during main grid fault contingencies. When the capacities are high ($\alpha>0.25$ and $\gamma>0.35$), further improving them yields a steep cost increase. The Pareto front provides a set of trade-off solutions from which designers can select the option that best meets their requirements.

\section{Conclusion}
\label{sec_conclusion}
This paper presented a scenario-based optimization framework to co-optimize BESS rated power and energy capacity for grid-interactive data centers while allocating capability across demand response, fast load fluctuation smoothing, and reactive power capacity for quasi-static voltage support during PCC voltage sags. Minute-level scheduling with SoC dynamics, ramp limits, and a controllable queue of delay-insensitive workloads ensures operational feasibility under combinations of demand response events and PCC voltage sags.

Case studies using a real data center load profile show that demand response primarily drives energy capacity needs, while voltage support drives substantially higher power rating due to reactive/apparent power requirements. When both services are required, the optimal design must satisfy the combined constraints, increasing total BESS cost. Stress tests further indicate that frequent demand response events during high-load periods can prevent sufficient recharge, reducing achievable load reduction guarantees. Overall, the proposed framework offers practical planning guidance by translating multi-service interconnection expectations into explicit BESS sizing and trade-offs, and it motivates future work with more detailed event modeling and system dynamics for smoothing and ride-through performance.

\bibliographystyle{IEEEtran}
\bibliography{reference} 

@techreport{shehabi2024datacenter,
  author       = {Shehabi, Arman and Newkirk, Andrew and Smith, Sarah and Hubbard, A. and Lei, N. and Siddik, M. and others},
  title        = {2024 {U}nited {S}tates Data Center Energy Usage Report},
  year         = {2024},
  institution  = {Lawrence Berkeley National Laboratory},
  number       = {LBNL-2001637},
  doi          = {10.71468/P1WC7Q},
  url          = {https://escholarship.org/uc/item/32d6m0d1},
}

@inproceedings{samsi2021supercloud,
  title={The {MIT} supercloud dataset},
  author={Samsi, Siddharth and Weiss, Matthew L and Bestor, David and Li, Baolin and Jones, Michael and Reuther, Albert and Edelman, Daniel and Arcand, William and Byun, Chansup and Holodnack, John and others},
  booktitle={2021 IEEE High Performance Extreme Computing Conference (HPEC)},
  pages={1--8},
  year={2021},
  organization={IEEE}
}

@online{pnnl_eset,
  title   = {{The Energy Storage Evaluation Tool (ESET™)}},
  author  = {{Pacific Northwest National Laboratory ESET Team}},
  url     = {https://eset.pnnl.gov/}
}

@online{datacenterknowledge2025ai,
  author       = {Shane Snider},
  title        = {Why {AI} Data Center Projects Face Years of Delays After Approval},
  year         = {2026},
  url          = {https://www.datacenterknowledge.com/energy-power-supply/why-ai-data-center-projects-face-years-of-delays-after-approval},
  urldate      = {2026-05-12}
}

@article{chen2020internet,
  title={Internet data centers participating in demand response: A comprehensive review},
  author={Chen, Min and Gao, Ciwei and Song, Meng and Chen, Songsong and Li, Dezhi and Liu, Qiang},
  journal={Renewable and Sustainable Energy Reviews},
  volume={117},
  pages={109466},
  year={2020},
  publisher={Elsevier}
}

@article{xie2025enhancing,
  title={Enhancing Data Center Low-Voltage Ride-Through},
  author={Xie, Yiheng and Cui, Wenqi and Wierman, Adam},
  journal={arXiv preprint arXiv:2510.03867},
  year={2025}
}

@article{takci2025data,
  title={Data centres as a source of flexibility for power systems},
  author={Takci, Mehmet T{\"u}rker and Qadrdan, Meysam and Summers, Jon and Gustafsson, Jonas},
  journal={Energy Reports},
  volume={13},
  pages={3661--3671},
  year={2025},
  publisher={Elsevier}
}

@article{dabbagh2017shaving,
  title={Shaving data center power demand peaks through energy storage and workload shifting control},
  author={Dabbagh, Mehiar and Hamdaoui, Bechir and Rayes, Ammar and Guizani, Mohsen},
  journal={IEEE Transactions on Cloud Computing},
  volume={7},
  number={4},
  pages={1095--1108},
  year={2017},
  publisher={IEEE}
}

@inproceedings{shi2016leveraging,
  title={Leveraging energy storage to optimize data center electricity cost in emerging power markets},
  author={Shi, Yuanyuan and Xu, Bolun and Zhang, Baosen and Wang, Di},
  booktitle={Proceedings of the Seventh International Conference on Future Energy Systems},
  pages={1--13},
  year={2016}
}

@article{cupelli2018data,
  title={Data center control strategy for participation in demand response programs},
  author={Cupelli, Lisette and Sch{\"u}tz, Thomas and Jahangiri, Pooyan and Fuchs, Marcus and Monti, Antonello and M{\"u}ller, Dirk},
  journal={IEEE Transactions on Industrial Informatics},
  volume={14},
  number={11},
  pages={5087--5099},
  year={2018},
  publisher={IEEE}
}

@article{azizi2026strengthening,
  title={Strengthening data center operations using grid-forming battery energy storage as a line-interactive uninterruptible power supply},
  author={Azizi, Arian and Morovati, Samaneh and Zamani, Amin and Piruzza, Joshua and Guo, David and Liu, Zhuoning and Taimela, Pasi},
  journal={International Journal of Electrical Power \& Energy Systems},
  volume={175},
  pages={111638},
  year={2026},
  publisher={Elsevier}
}

@inproceedings{cupelli2017optimal,
  title={Optimal sizing of data center battery energy storage system for provision of frequency containment reserve},
  author={Cupelli, Lisette and Barve, Nikhil and Monti, Antonello},
  booktitle={IECON 2017-43rd Annual Conference of the IEEE Industrial Electronics Society},
  pages={7185--7190},
  year={2017},
  organization={IEEE}
}

@article{liu2026watts,
  title={{W}atts vs. {B}ytes: {T}urning Data Centers into Grid Assets via Storage---Compute Co-Optimization},
  author={Liu, Shaohui and Shin, Sungho and Deka, Deepjyoti},
  journal={arXiv preprint arXiv:2605.16190},
  year={2026},
  url={https://arxiv.org/abs/2605.16190},
  note={Submitted to arXiv}
}

@article{thompson2016optimization,
  title={Optimization of data center battery storage investments for microgrid cost savings, emissions reduction, and reliability enhancement},
  author={Thompson, Christopher C and Oikonomou, PE Konstantinos and Etemadi, Amir H and Sorger, Volker J},
  journal={IEEE Transactions on Industry Applications},
  volume={52},
  number={3},
  pages={2053--2060},
  year={2016},
  publisher={IEEE}
}

@article{mamun2016multi,
  title={Multi-objective optimization of demand response in a datacenter with lithium-ion battery storage},
  author={Mamun, A and Narayanan, I and Wang, D and Sivasubramaniam, A and Fathy, HK},
  journal={Journal of Energy Storage},
  volume={7},
  pages={258--269},
  year={2016},
  publisher={Elsevier}
}

\end{document}